\documentclass[11pt,a4paper]{article}
\usepackage{geometry}
\usepackage{graphics}
\usepackage{setspace}
\usepackage[scaled=0.9]{helvet}
\usepackage{graphicx}
\usepackage{subfig}
\usepackage{amssymb}
\usepackage{rotating}
\usepackage{arydshln}
\usepackage{multicol}
\usepackage{abstract}
\usepackage{ucs}
\usepackage[utf8x,utf8]{inputenc}
\usepackage[T1]{fontenc}
\usepackage[swedish, english]{babel}
\usepackage{fancyhdr}

\begin{document}
\bibliographystyle{acm}
\pagestyle{fancy}
\cfoot{\thepage}
\renewcommand{\abstractname}{}

\title{\fontfamily{phv}\selectfont{\huge{\bfseries{Data-driven estimation of neutral pileup particle multiplicity in high-luminosity hadron collider environments}}}}
\author{
{\fontfamily{ptm}\selectfont{\large{Federico Colecchia}}}\thanks{Email: federico.colecchia@brunel.ac.uk}\\
{\fontfamily{ptm}\selectfont{\large{{\it Brunel University London, Kingston Lane, Uxbridge UB8 3PH, UNITED KINGDOM}}}}
}
\date{}
\maketitle
\begin{onecolabstract}
The upcoming operation regimes of the Large Hadron Collider are going to place stronger requirements on the rejection of particles originating from pileup, i.e.\ from interactions between other protons. For this reason, particle weighting techniques have recently been proposed in order to subtract pileup at the level of individual particles. We describe a choice of weights that, unlike others that rely on particle proximity, exploits the particle-level kinematic signatures of the high-energy scattering and of the pileup interactions. We illustrate the use of the weights to estimate the number density of neutral pileup particles inside individual events, and we elaborate on the complementarity between ours and other methods. We conclude by suggesting the idea of combining different sets of weights with a view to exploiting different features of the underlying processes for improved pileup subtraction at higher luminosity.
\end{onecolabstract}

\begin{multicols}{2}

\section{Introduction\label{intro}}

The contamination, or background, from low-energy processes described by Quantum Chromodynamics (QCD) is a major challenge at the Large Hadron Collider (LHC), and the potential impact on physics analysis is anticipated to become even more significant in the upcoming operation regimes of the accelerator. In fact, the average pileup rate, i.e.\ the rate of low-energy interactions between other protons, will increase with the instantaneous luminosity of the collider, and this is anticipated to place stronger requirements on the correction techniques employed at the LHC experiments.

The presence of multiple vertices inside collision events due to pileup can significantly complicate the extraction from the data of the physics quantities of interest, and calls for the use of dedicated subtraction techniques. Established methods that are part of the core reconstruction pipelines at the LHC experiments rely on the use of tracking information for charged particles, as well as on estimates of the pileup energy flow associated with neutral\footnote{
Whenever neutral particles are referred to in the text, neutrinos are not considered.
} particles \cite{REVIEW_PILEUP_2014}, for which the task of assigning a vertex of origin is in general significantly more difficult.

In the light of the upcoming operation regimes of the LHC, algorithms of a different nature have recently been proposed and are currently being evaluated. Methods such as those presented in \cite{PUPPI,SoftKiller,berta} assign individual particles inside collision events weights that reflect the probability of the particles originating from soft, i.e.\ low-energy, QCD interactions as opposed to the signal hard parton scattering. The use of the weights to rescale the particle four-momentum vectors \cite{PUPPI} has been shown to result in improved performance in terms of pileup subtraction when compared to traditional methods. It is worth noting that the implications of assigning weights to individual particles were also explored from a different perspective in \cite{jet-sampling}, where a connection was established with the idea of multiple interpretations of the data, and where the potential benefit to physics analysis was discussed.

\section{Our approach{\label{method}}}


\begin{figure*}
\centering
\subfigure{
\includegraphics[scale=0.43]{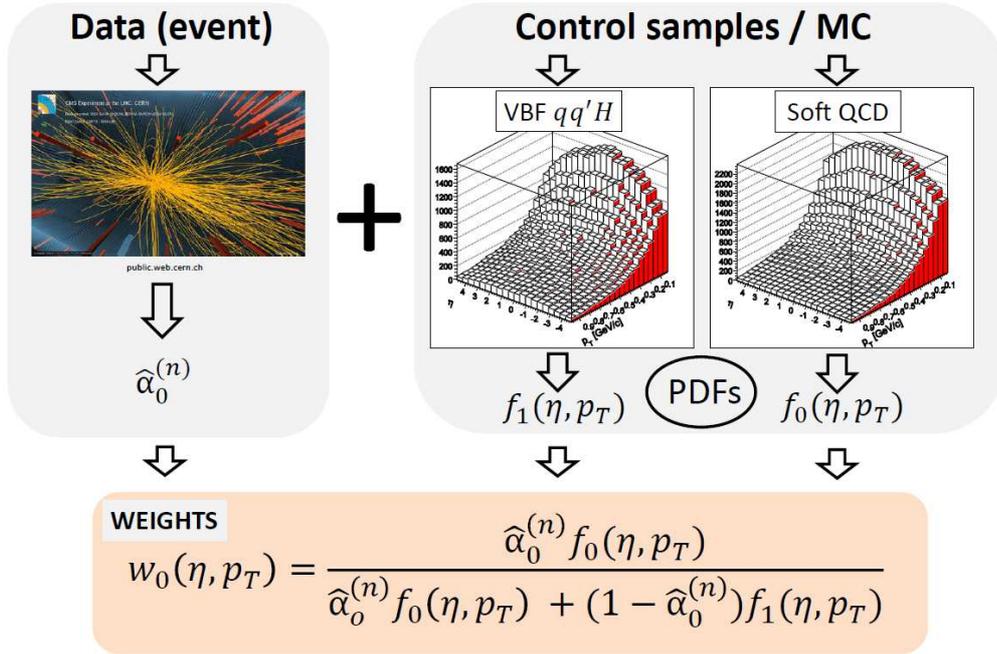}
}
\caption{
Schematic representation of the data processing involved in the calculation of the particle weights, as described in the text. High-statistics control samples are used to estimate the shapes of the particle-level PDFs for signal (Standard Model Higgs boson production via vector boson fusion (VBF) from proton-proton interactions at $\sqrt{s}=14~$TeV) and for background soft QCD interactions. The overall fraction of neutral particles originating from background is calculated event by event as described in section \ref{fractions}. This information is then combined into the particle weights, $w_0(\eta,p_T)$, which reflect the probability of individual neutral particles originating from soft QCD interactions as opposed to the hard parton scattering, based on the $(\eta, p_T)$ bin they belong to. The results reported in this article correspond to 50 pileup vertices per event.
}
\label{fig:overview_weights}
\end{figure*}

The contribution of this article is two-fold. 

\begin{itemize}
\item {\it A different choice of particle weights}. We are proposing a different definition of the particle weights, which makes use of information about the underlying physics that, to our knowledge, is not employed by other methods. As opposed to exploiting the existence of collinear singularities in the physics that underlies the showering process \cite{PUPPI}, our technique relies on the particle-level kinematic signatures of the hard parton scattering and of the soft QCD interactions. Our aim is to estimate the variability in the shapes of the distributions inside individual collision events that is associated with statistical fluctuations in the data.
\item {\it A different application of the weights}. We investigate a different use of the weights, whereby the particle-level kinematic distributions in the data are reshaped in order to estimate the number of neutral pileup particles in different kinematic regions inside individual events. 
\end{itemize}

We treat each event as a heterogeneous statistical population of particles that have their origin either in the signal hard parton scattering or in background soft QCD interactions. Although it is generally not possible to map individual particles to a single physics process in a hadron collider environment due to colour connection, the use of weights that reflect the likelihood of particles originating from either process provides a way of addressing this conceptual issue.  

Figure \ref{fig:overview_weights} shows a schematic representation of the procedure employed to calculate the weights. The shapes of the particle-level probability density functions (PDFs) corresponding to the signal high-energy parton scattering (in this article, Standard Model Higgs boson production via vector boson fusion) and to soft QCD interactions are obtained from control samples. Since the data sets in question are high-statistics, the effect of statistical fluctuations in the data is averaged out, and the shapes of the distributions reflect the expectation from the underlying physics processes. On the other hand, the overall fraction of neutral particles associated with soft QCD interactions is estimated event by event, which takes into account the variability of the neutral pileup particle fraction across collisions. This information is then combined into the particle weights, $w_0(\eta, p_T)$, which reflect the expected fraction of neutral soft QCD particles in each event as a function of particle $\eta$ and $p_T$.
%

We discuss the results of an initial study on Monte Carlo data at the generator level, and show that our weights can be used to estimate the number of neutral pileup particles across the kinematic space inside individual events with reasonable accuracy. 




The technique described in this article is based on a deterministic variant of a Markov Chain Monte Carlo algorithm that we proposed for particle-by-particle filtering of individual events at the reconstruction level \cite{gibbshep,gibbshep2}. Our main goal is to contribute to improve on the subtraction of soft QCD background in high-luminosity hadron collider environments using algorithms that can be implemented at the reconstruction level. Specifically, we are targetting a processing stage upstream of jet reconstruction, i.e.\ before the particles are clustered together according to their likelihood of originating from the same scattered hard parton. The advantages of this algorithm over the previous stochastic version are its parallel nature and the simplicity of the calculations involved. 

The results shown in this article complement those discussed in \cite{count_arxiv} with reference to a different signal process: as opposed to $t\bar{t}$ production via gluon fusion, the signal distributions reported in the following relate to Standard Model Higgs boson production via vector boson fusion, which does not involve colour exchange between the colliding protons.






\section{The algorithm\label{algo}}

This section describes the calculation of the particle weights, as well as the use of the weights to estimate the number of neutral pileup particles in different regions of the particle $(\eta, p_T)$ space inside individual events. This study concentrates on the region $-5<\eta<5$, $0<p_T<1$~GeV/c, which contains most particles associated with soft QCD interactions. The $(\eta, p_T)$ space was subdivided into bins of widths $\Delta\eta = 0.5$ and $\Delta p_T = 0.05$~GeV/c. 

\subsection{Control sample PDF templates\label{cs}}

High-statistics control samples were used to obtain the shapes of the average particle-level $(\eta, p_T)$ distributions corresponding to particles originating from the signal hard parton scattering and from background soft QCD interactions. Monte Carlo data sets were generated using Pythia 8.176 \cite{pythia1,pythia2}, corresponding to the following:

\begin{itemize}
\item {\bf\em Signal}: $\sim$300,000 final-state particles associated with Standard Model Higgs boson production via vector boson fusion, i.e.\ $qq^{\prime}\rightarrow qq^{\prime}WW(ZZ)\rightarrow qq^{\prime}H$ , from $pp$ collisions at $\sqrt{s}=14$~TeV.
\item {\bf\em Background}: $\sim$300,000 soft QCD particles corresponding to 50 pileup vertices per event.
\end{itemize}

Such high-statistics distributions reflect the expectation from the corresponding physics processes whereby the effect of statistical fluctuations in the data is averaged out, and can therefore be used to estimate the expected, or average, number of neutral pileup particles in each $(\eta, p_T)$ bin. On the other hand, the corresponding unknown actual numbers generally deviate from the expected values, and, given the typical particle multiplicity inside LHC events, the discrepancy is often non-negligible. The calculation of the statistical uncertainty on the estimated number of neutral soft QCD particles in each $(\eta, p_T)$ bin is discussed in section \ref{fingerprints}. 

The high-statistics $(\eta, p_T)$ distributions obtained in this study for neutral final-state particles associated with the hard parton scattering and with soft QCD interactions are shown as part of the schematic representation of the workflow of the analysis in figure \ref{fig:overview_weights}. The plots were rotated around the vertical axis in order to make the distributions more clearly visible.



\subsection{Event-by-event neutral particle fractions \label{fractions}}

In addition to the shapes of the probability distributions described in section \ref{cs}, the calculation of the expected number of neutral soft QCD particles in each $(\eta, p_T)$ bin, $\nu_b(\eta, p_T)$, also requires an event-by-event estimate of the overall fraction of neutral particles originating from soft QCD interactions as opposed to the hard parton scattering, $\hat{\alpha}_0^{(n)}$. For the purpose of calculating the particle weights, assigning a value to $\hat{\alpha}_0^{(n)}$ is essentially equivalent to specifying the relative normalisation of the signal and background distributions.

The quantity $\hat{\alpha}^{(n)}$ was estimated in each event based on the corresponding charged particle fraction, $\hat{\alpha}_0^{(c)}$, according to this formula: 

\begin{equation}
\hat{\alpha}_0^{(n)}=\mbox{min}(k \hat{\alpha}_0^{(c)}, \hat{\alpha}_0^{(c)}),
\label{eq:alpha_0}
\end{equation}

where the use of ``min'' ensures that $\hat{\alpha}_0^{(n)}\in [0,1]$.

\begin{figure*}
\centering
\begin{minipage}{17pc}
\includegraphics[scale=0.3]{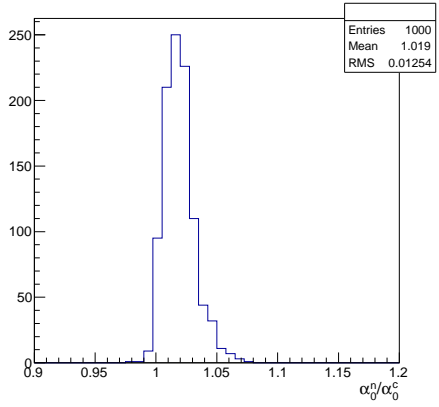}
\end{minipage}\hspace{3pc}%
\begin{minipage}{12pc}
\caption{\label{fig:fracs}Ratio between the fraction of neutral particles associated with soft QCD interactions and the corresponding charged fraction, $\alpha_0^{(n)}/\alpha_0^{(c)}$, from Monte Carlo. The distribution was obtained over the events generated in this study.}
\end{minipage}
\end{figure*}

The correction factor $k$ takes into account the difference between charged and neutral particle kinematics, including the effect of charged particles with $p_T \lesssim 0.5$~GeV/c not reaching the tracking detectors. The ratio between the fraction of neutral soft QCD particles and the corresponding charged fraction, calculated using Monte Carlo, is shown in figure \ref{fig:fracs}. The distribution was obtained over the events analysed in this study, and the corresponding average value, $\left<k\right>=1.02$, was used as the value of $k$.

\begin{figure*}
\centering
\subfigure[]{
\includegraphics[scale=0.38]{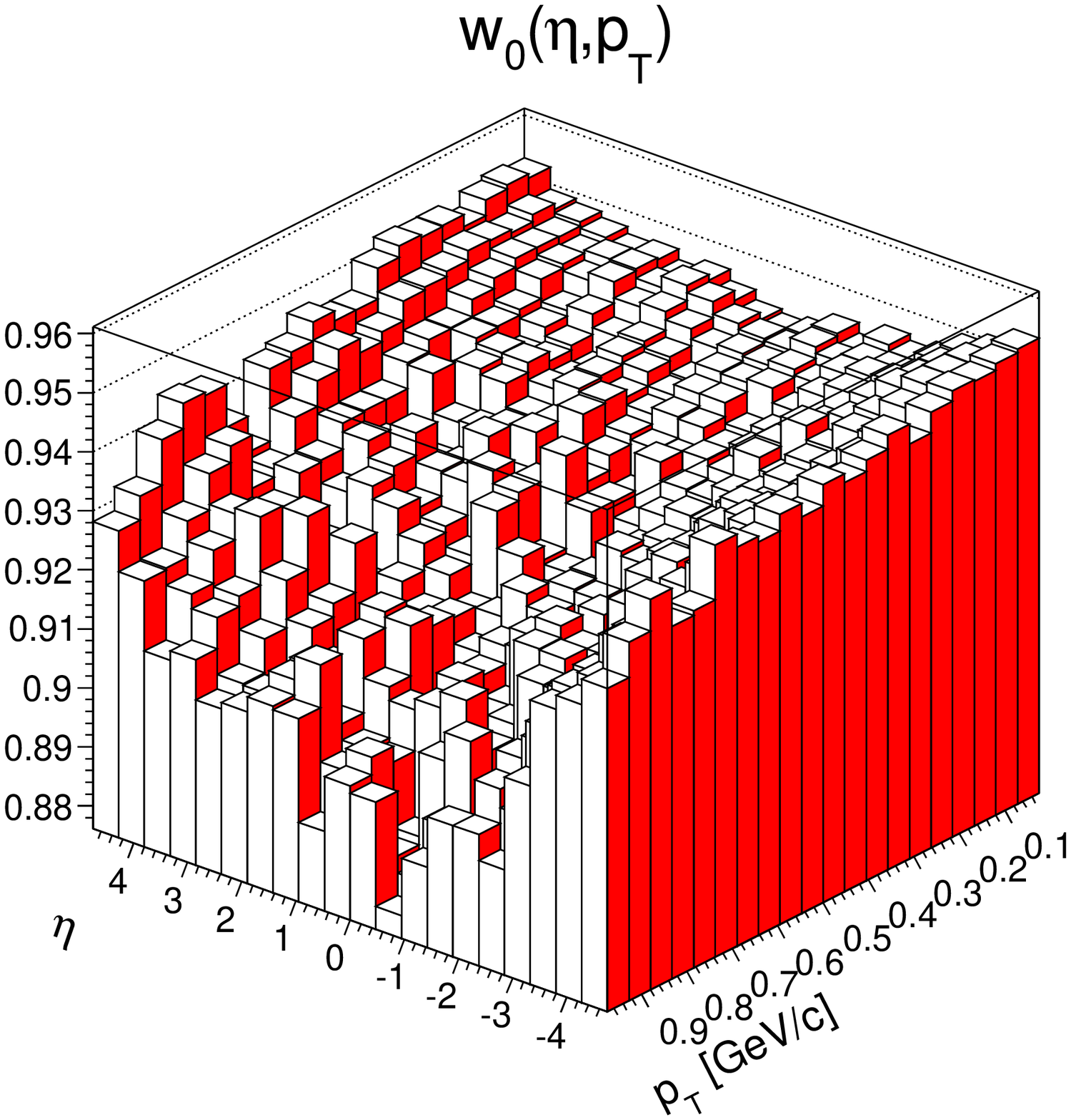}
}
\subfigure[]{
\includegraphics[scale=0.38]{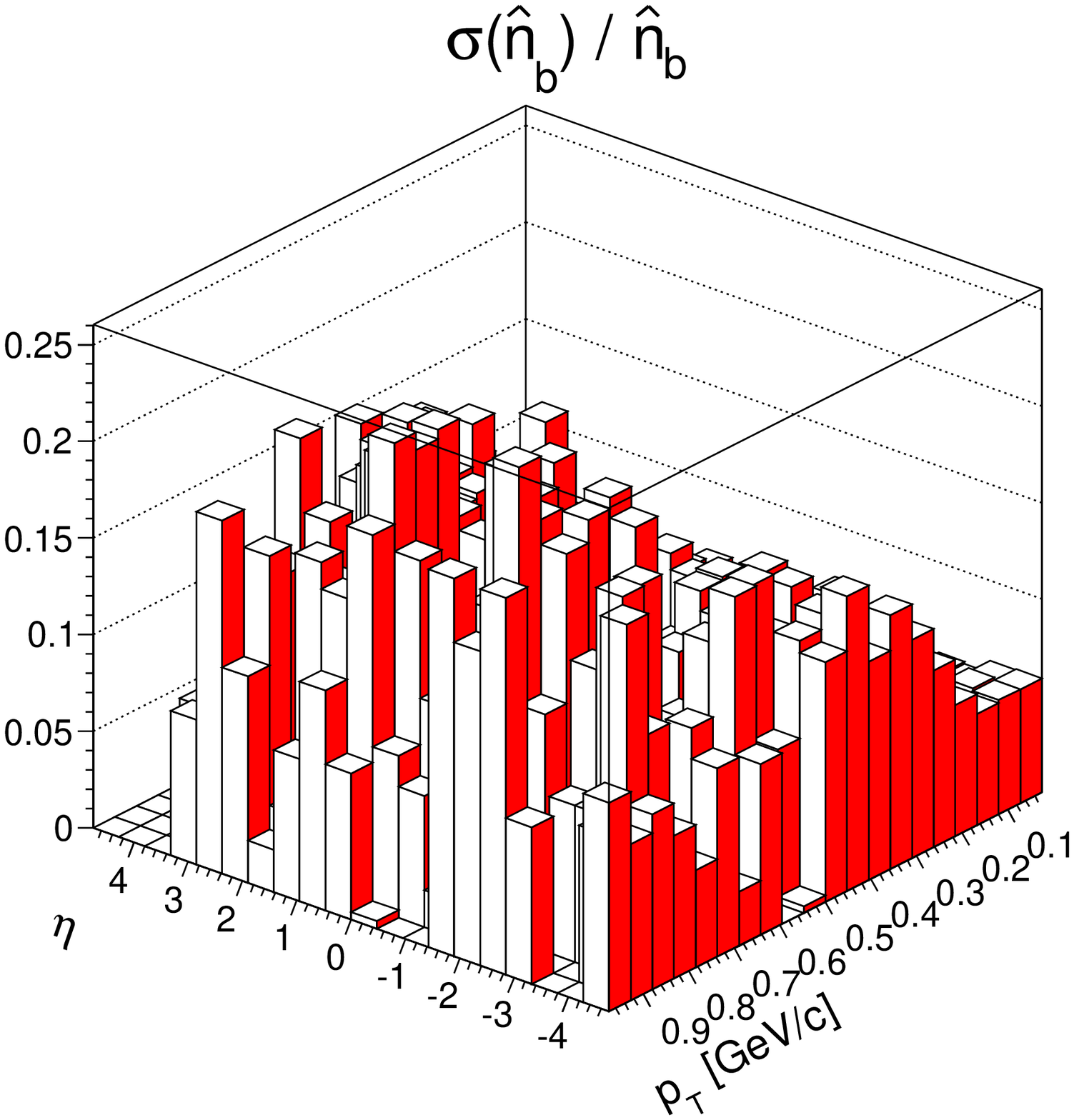}
}
\caption[]{
(a) Particle weights, $w_0(\eta, p_T)$, across the $(\eta, p_T)$ space in the region investigated, $-5<\eta<5$, $0<p_T<1$~GeV/c. The plot, which corresponds to one of the events analysed, was rotated around the vertical axis in order to make the distribution more clearly visible. (b) Relative uncertainty on the estimated number of neutral pileup particles, $\sigma_{\hat{n}_b}(\eta, p_T)/\hat{n}_b(\eta, p_T)$.}
\label{fig:heatmap}
\end{figure*}

\begin{figure*}
\centering
\subfigure[]{
\includegraphics[scale=0.38]{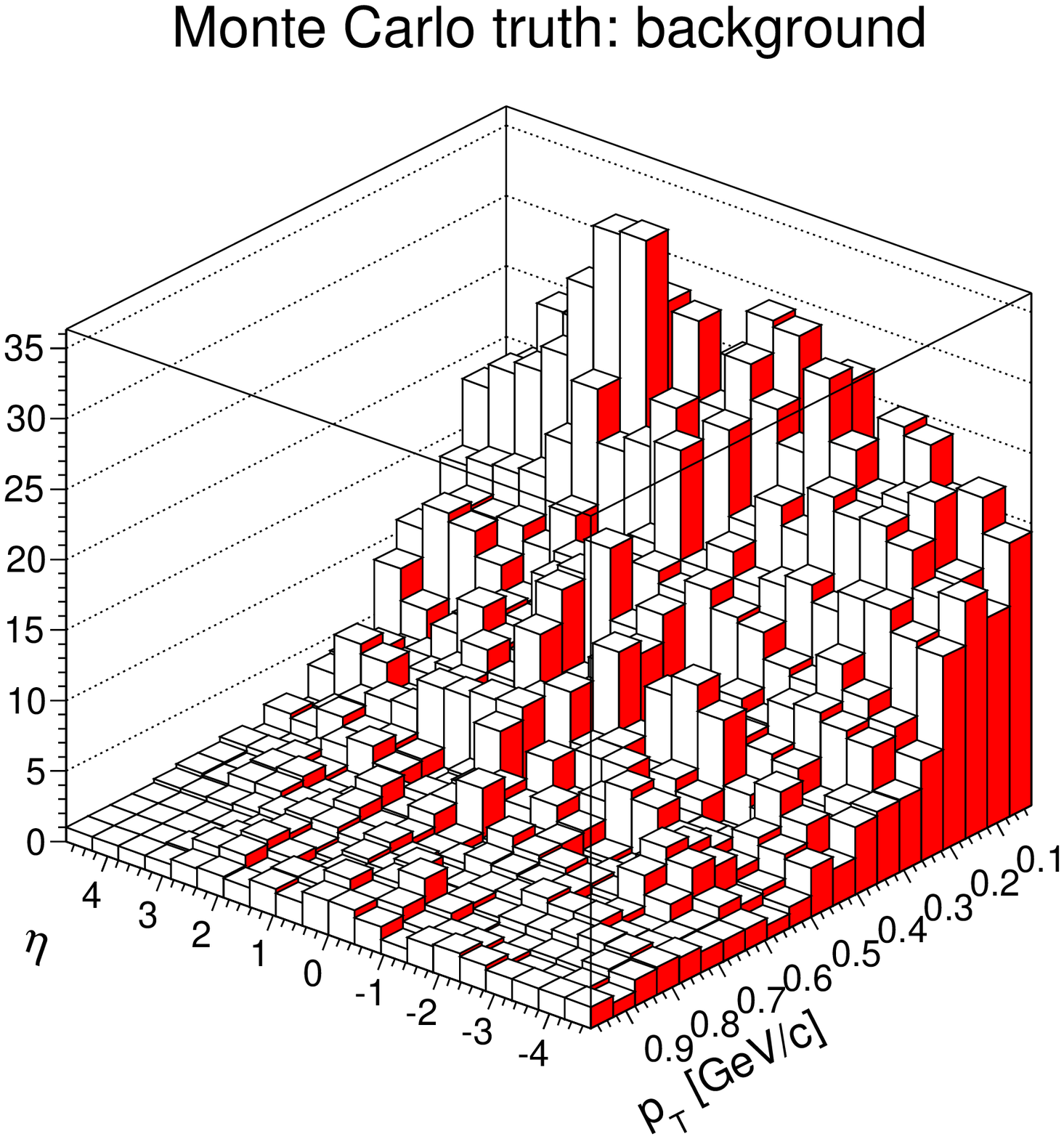}
}
\subfigure[]{
\includegraphics[scale=0.38]{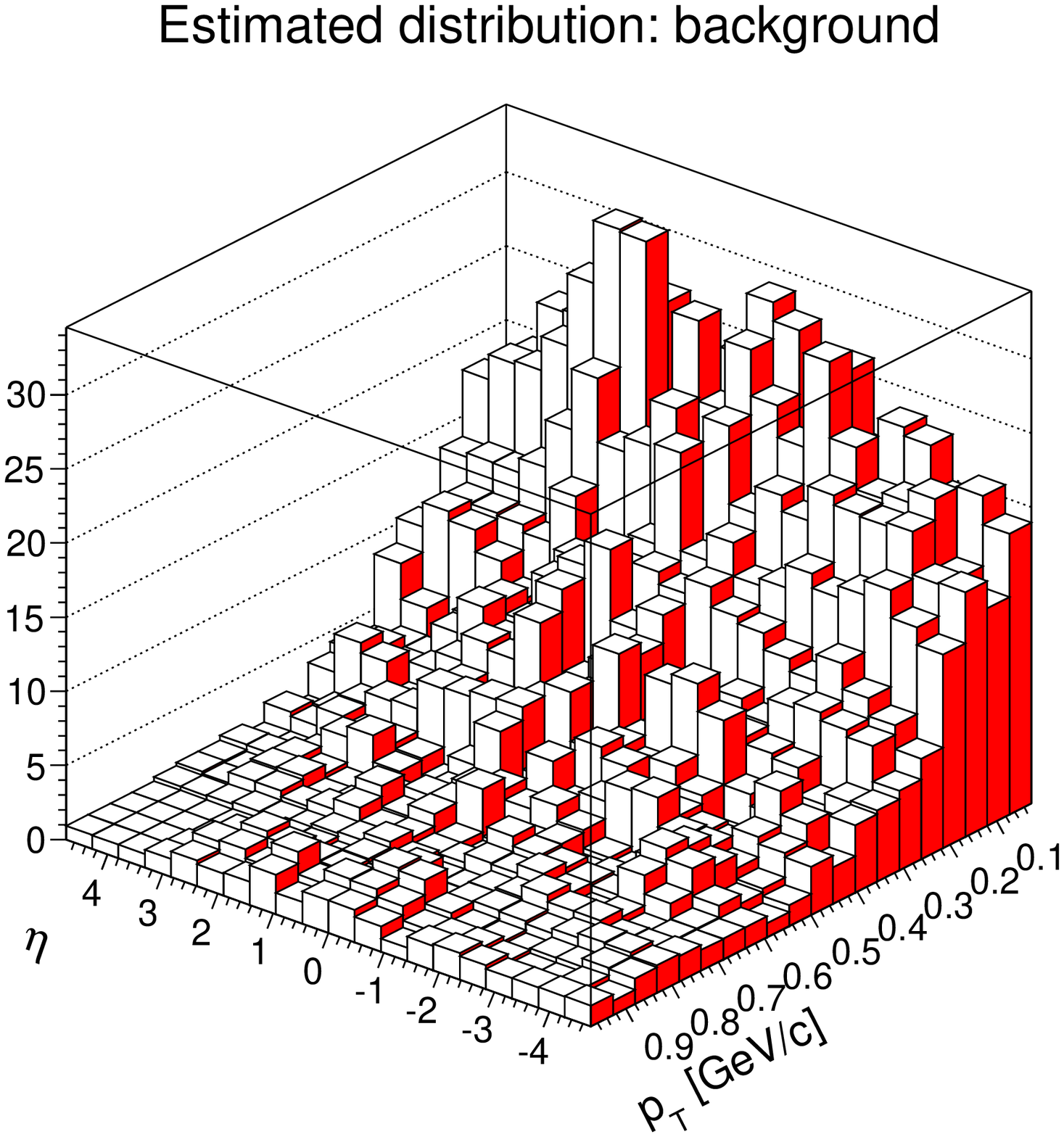}
}
\caption[]{
(a) True particle-level $(\eta, p_T)$ distribution of neutral soft QCD particles corresponding to one of the events analysed in this study. The effect of statistical fluctuations on the shape of the distribution is apparent. (b) The corresponding $(\eta, p_T)$ distribution estimated using this algorithm.}
\label{fig:truth}
\end{figure*}

\begin{figure*}
\centering
\subfigure[]{
\includegraphics[scale=0.17]{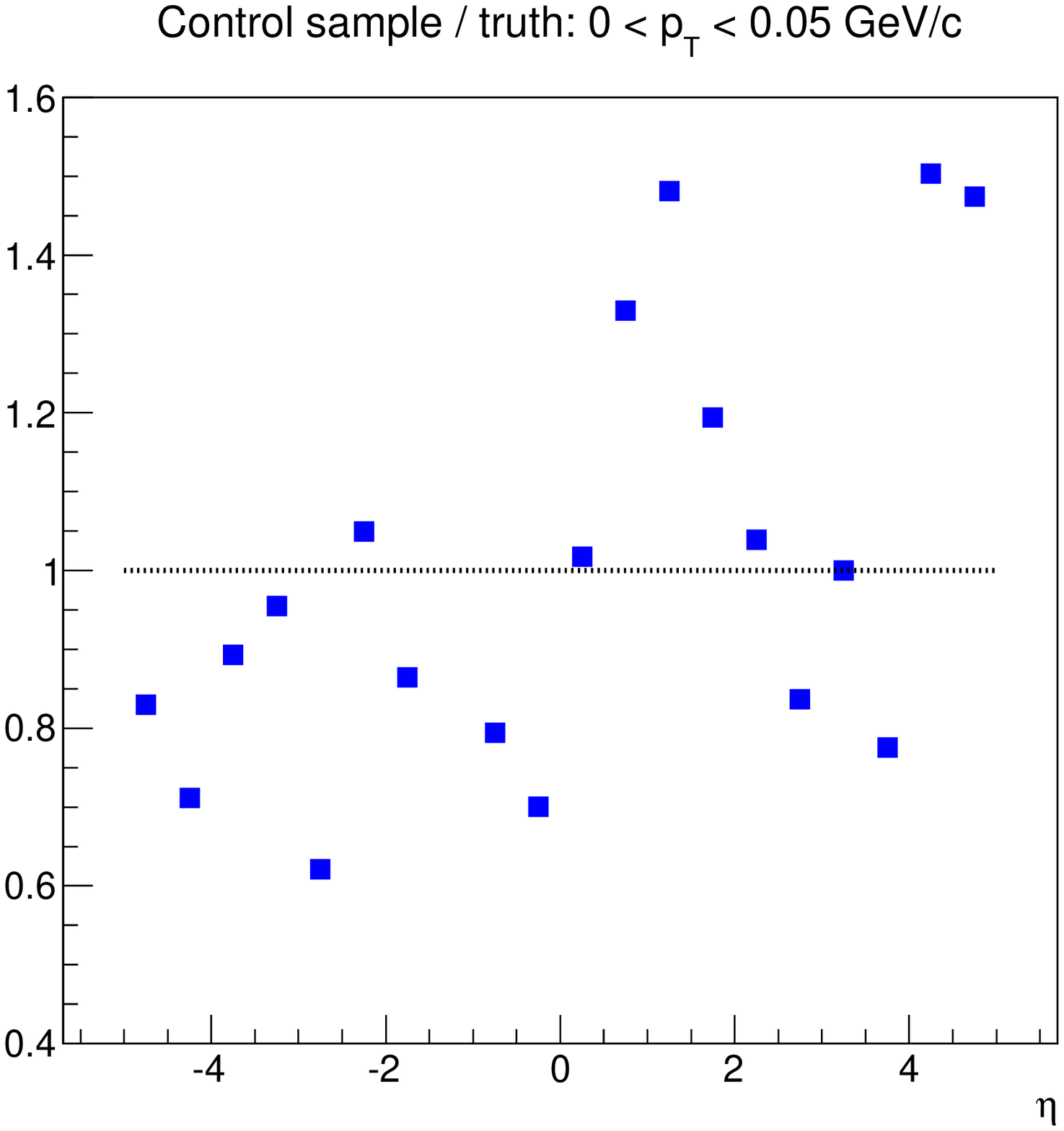}
}
\subfigure[]{
\includegraphics[scale=0.17]{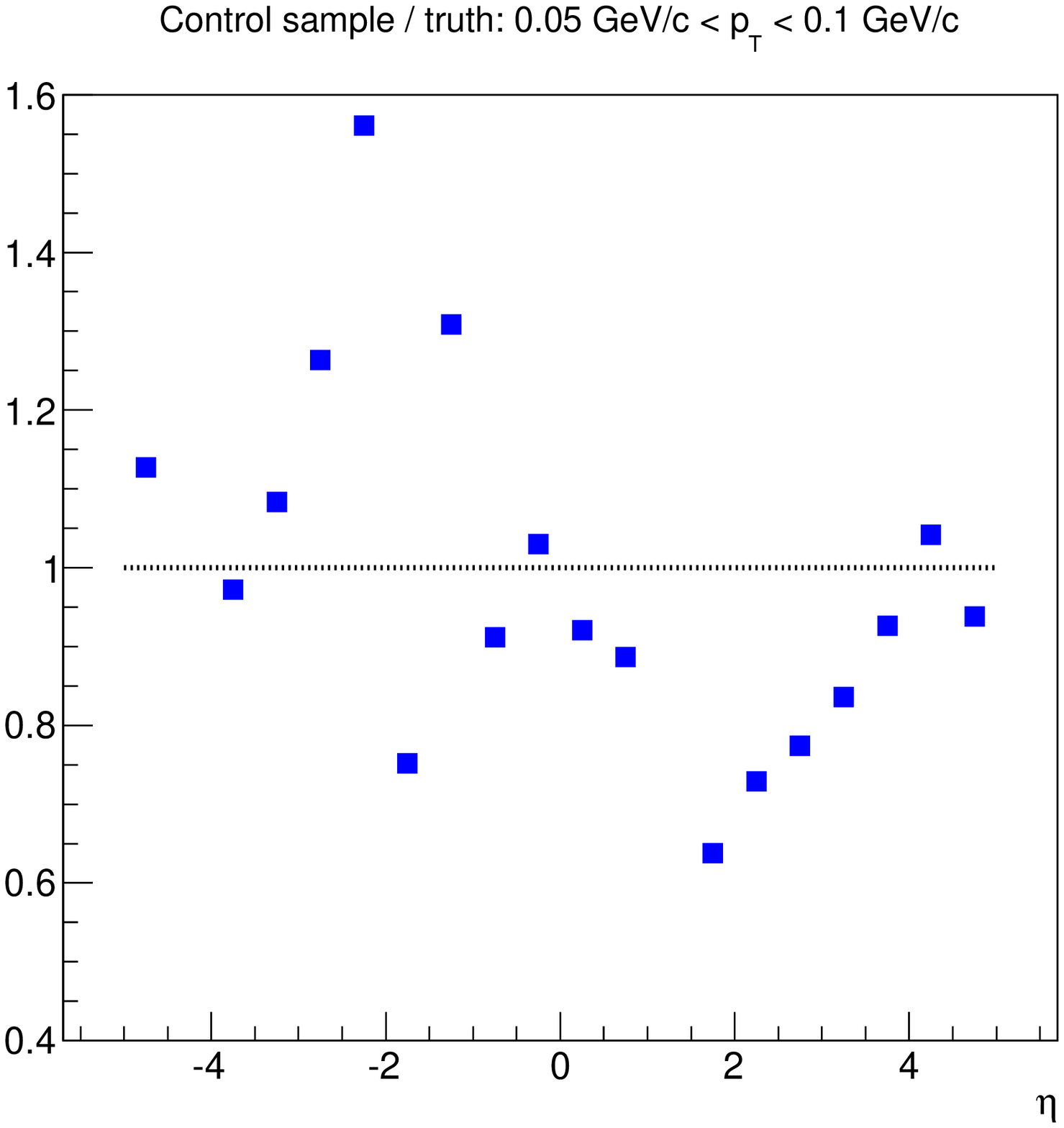}
}
\subfigure[]{
\includegraphics[scale=0.17]{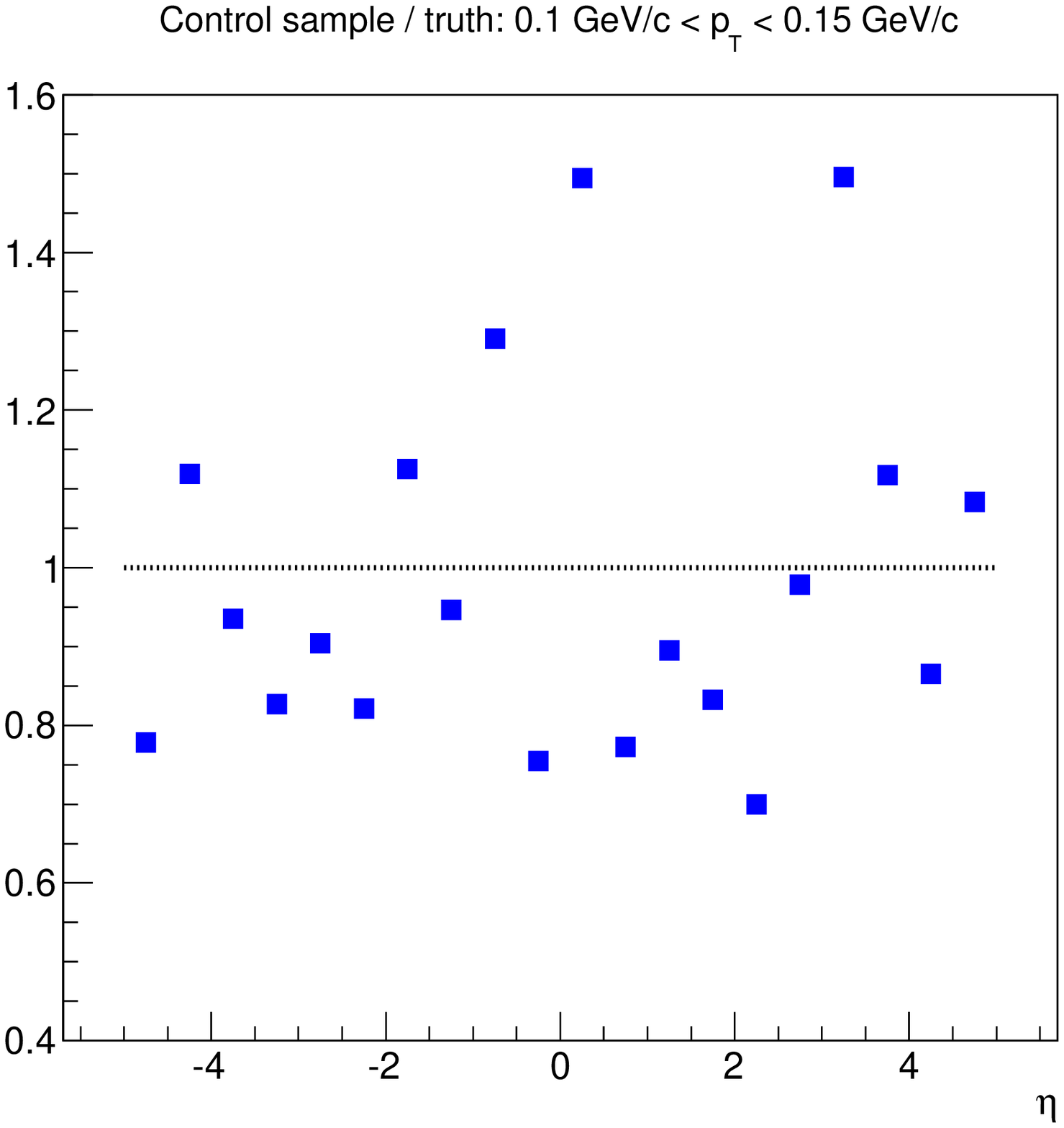}
}
\subfigure[]{
\includegraphics[scale=0.17]{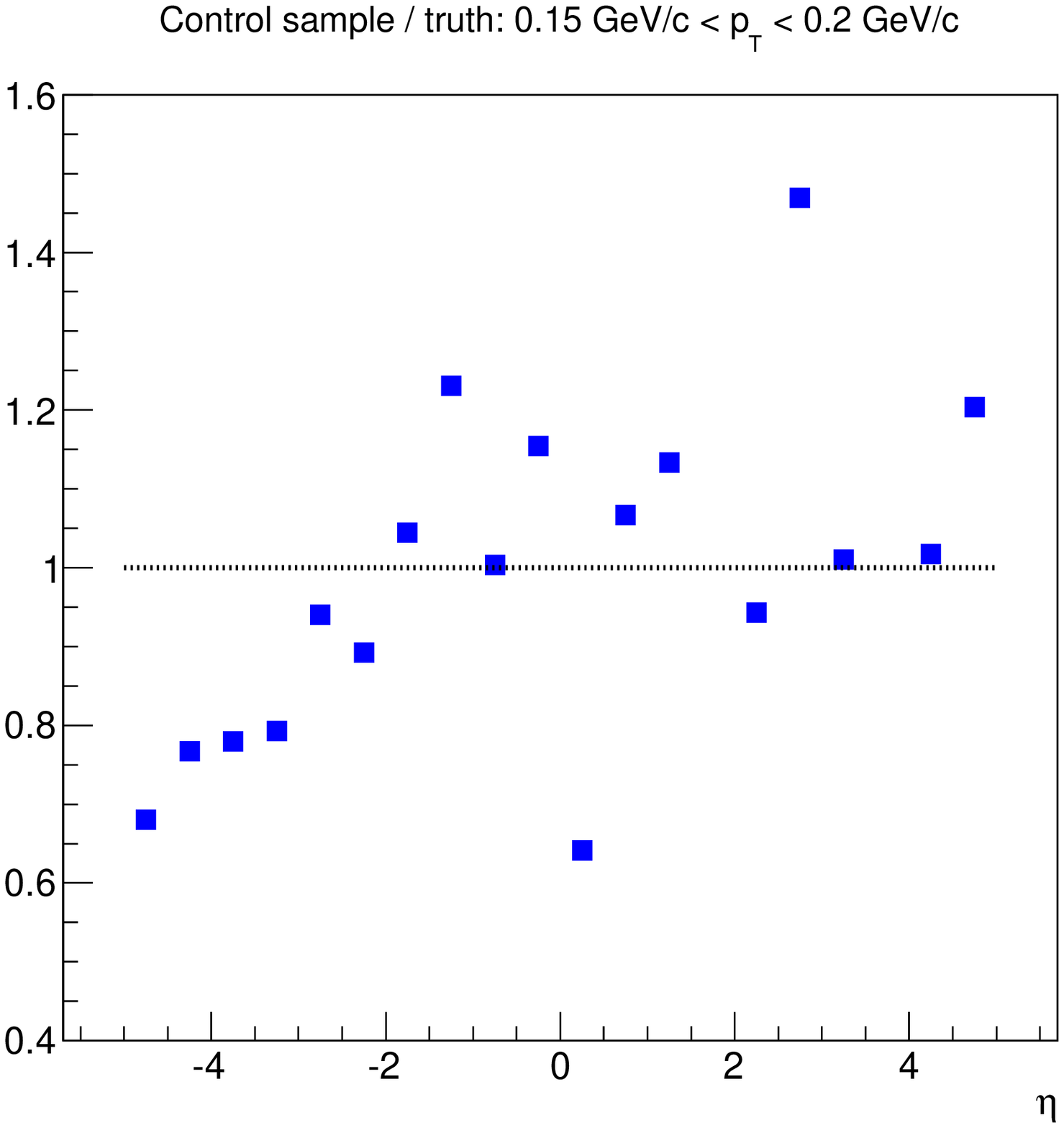}
}\\
\subfigure[]{
\includegraphics[scale=0.17]{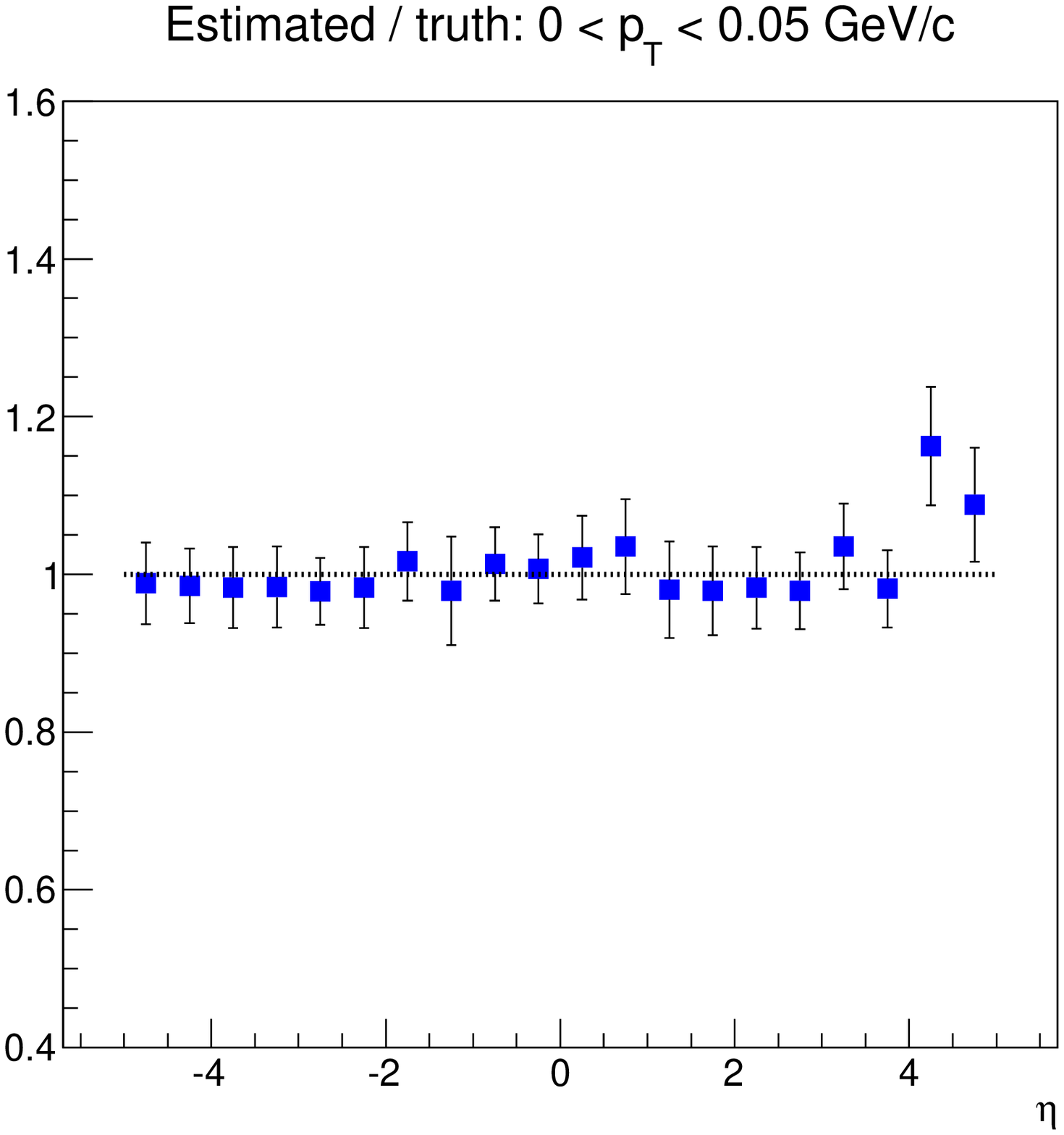}
}
\subfigure[]{
\includegraphics[scale=0.17]{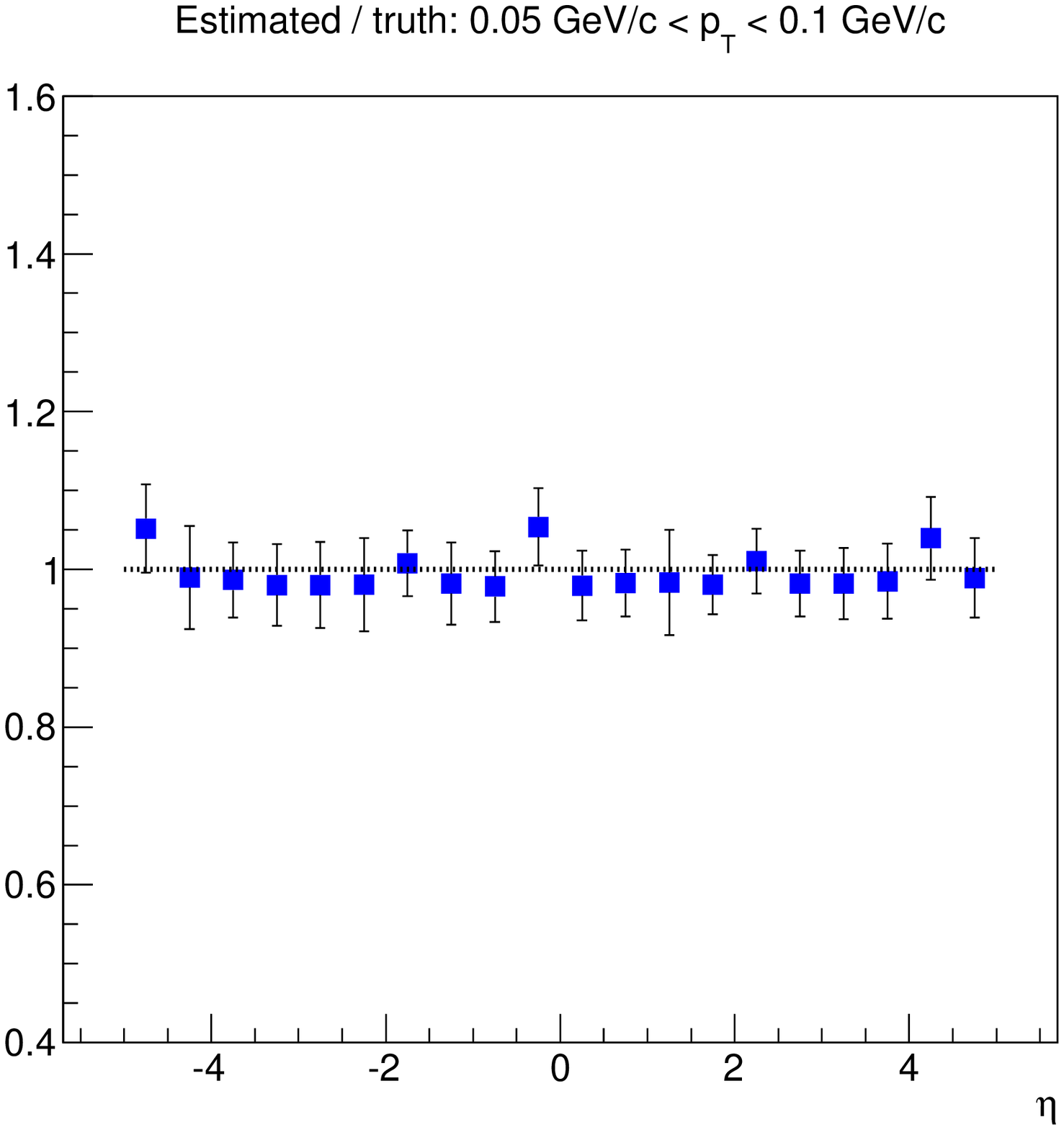}
}
\subfigure[]{
\includegraphics[scale=0.17]{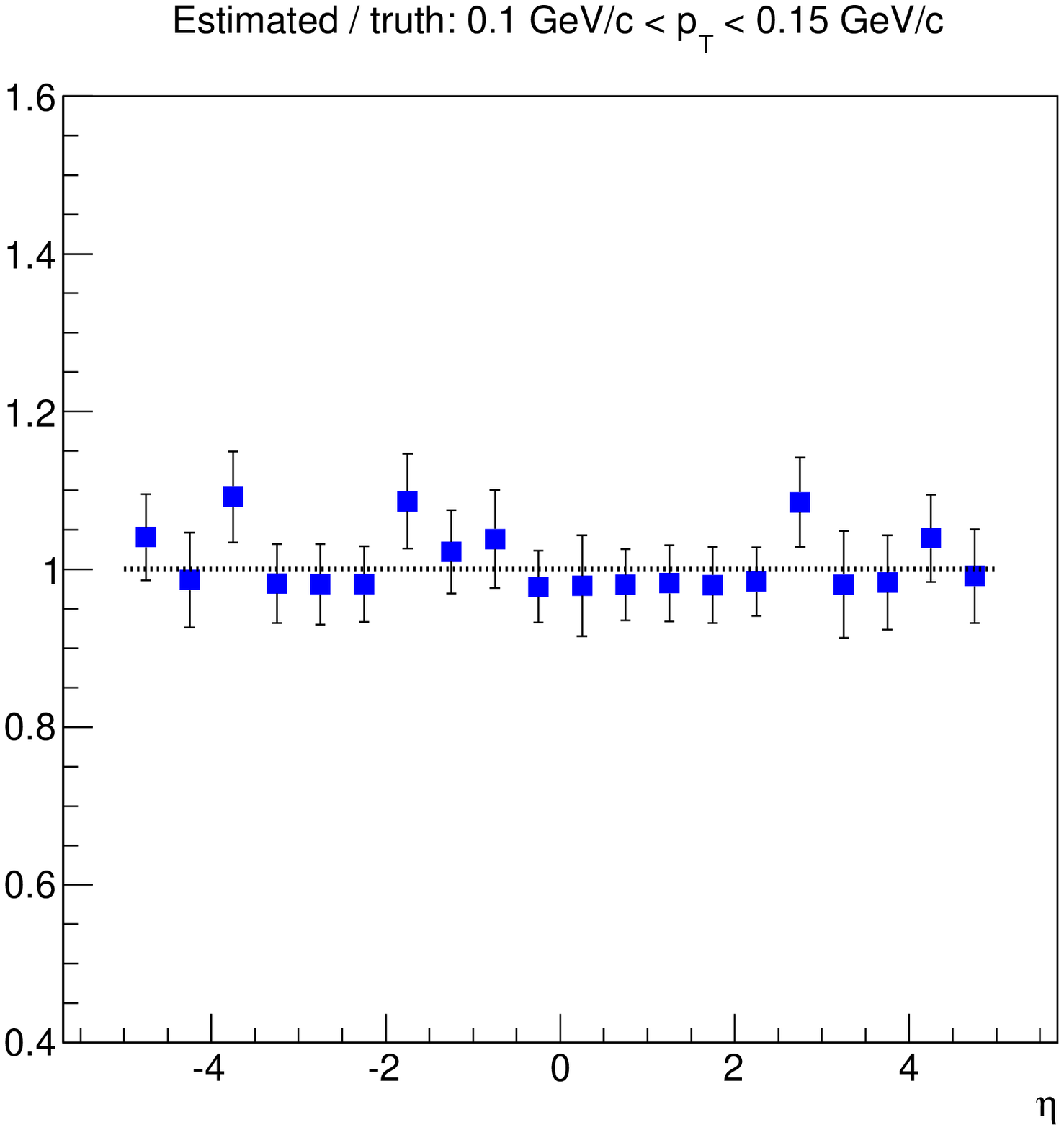}
}
\subfigure[]{
\includegraphics[scale=0.17]{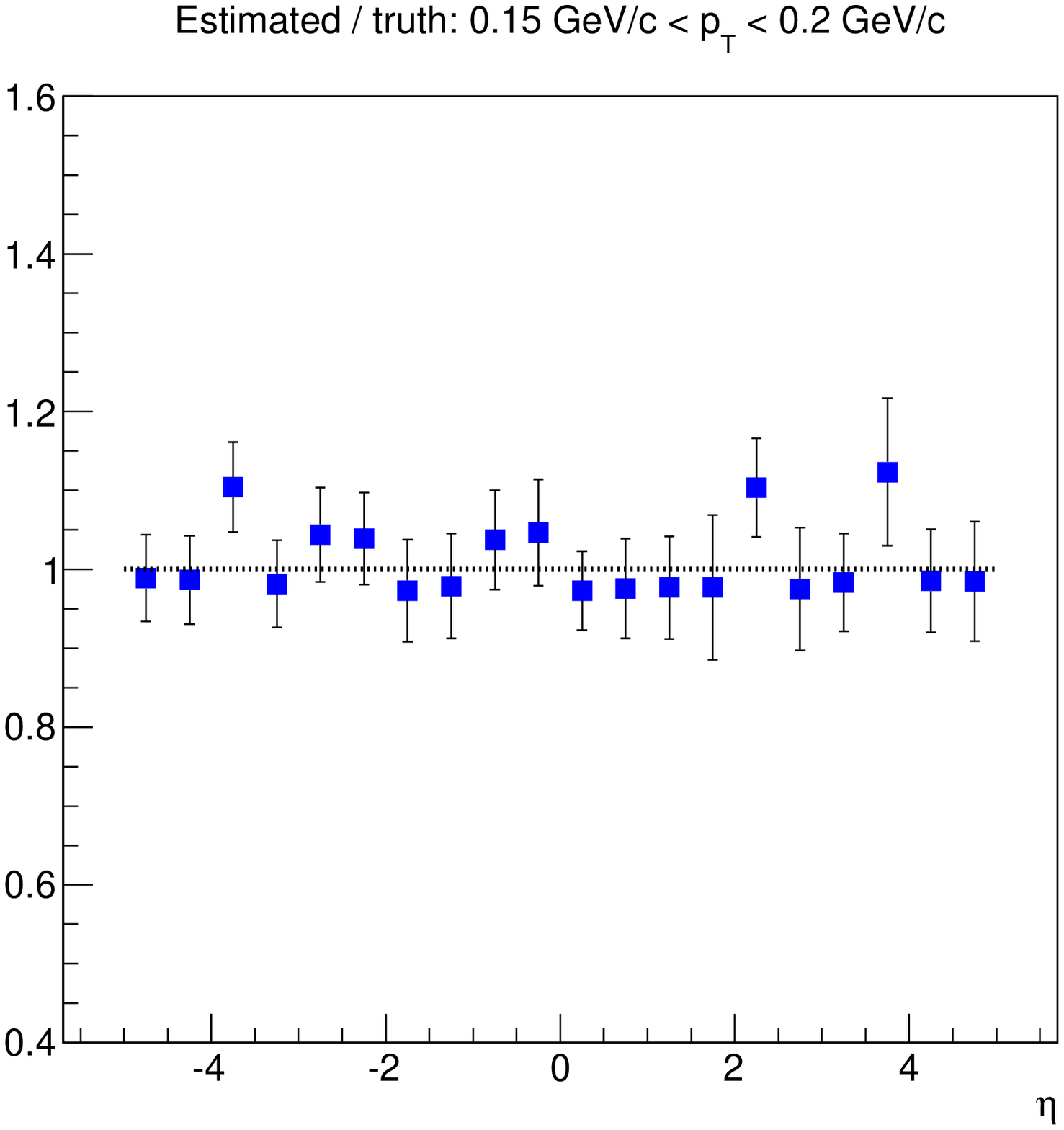}
}
\caption[]{
(a-d) Ratio between the control sample and the true $(\eta, p_T)$ distribution of neutral soft QCD particles, both normalised to unit volume. The ratios are shown as a function of particle $\eta$ in different $p_T$ bins in the region $0<p_T<0.2$~GeV/c. The error bars (not visible on this scale) correspond to one Poisson standard deviation on the control sample bin contents. (a) $0<p_T<0.05$~GeV/c. (b) $0.05~\mbox{GeV/c}<p_T<0.1$~GeV/c. (c) $0.1~\mbox{GeV/c}<p_T<0.15$~GeV/c. (d) $0.15~\mbox{GeV/c}<p_T<0.2$~GeV/c. (e-h) Ratio between the estimated $(\eta, p_T)$ distribution of neutral soft QCD particles and the corresponding true distribution, both normalised to unit volume. The ratio is displayed as a function of particle $\eta$ in different $p_T$ bins, and the error bars correspond to one binomial standard deviation on the bin contents in $\hat{n}_b(\eta, p_T)$. (e) $0<p_T<0.05$~GeV/c. (f) $0.05~\mbox{GeV/c}<p_T<0.1$~GeV/c. (g) $0.1~\mbox{GeV/c}<p_T<0.15$~GeV/c. (h) $0.15~\mbox{GeV/c}<p_T<0.2$~GeV/c.
}
\label{fig:rats}
\end{figure*}

\subsection{Particle weights\label{weights}}

The above information was combined into the definition of the particle weights as follows:

\begin{equation}
w_0(\eta, p_T) = \displaystyle \frac{\hat{\alpha}_0^{(n)} f_0(\eta, p_T)}{\hat{\alpha}_0^{(n)} f_0(\eta, p_T) + \hat{\alpha}_1^{(n)} f_1(\eta, p_T)},
\label{eq:w_0}
\end{equation}

where $\hat{\alpha}_1^{(n)} \equiv 1-\hat{\alpha}_0^{(n)}$. Figure \ref{fig:heatmap} (a) displays $w_0(\eta, p_T)$ in the region $-5<\eta<5$ and $0<p_T<1$~GeV/c, corresponding to one of the events analysed. 

This choice of weights reflects the probability of individual particles originating from soft QCD interactions as opposed to the hard parton scattering, based on the shapes of the expected $(\eta, p_T)$ PDFs as well as on the estimated overall fraction of neutral soft QCD particles in each event. This highlights the difference between our approach and other weighting methods that rely on particle-to-particle proximity measures and that exploit different properties of the underlying physics, e.g.\ the existence of collinear singularities in the showering process \cite{PUPPI}.

It should be noted that our decomposition of the particle $(\eta, p_T)$ space is relatively coarse-grained, particularly along the $\eta$ axis. This is one of the reasons why we are not proposing that these weights be used in isolation, but rather in conjunction with those employed by other methods. It is our opinion that combining different sets of weights reflecting different properties of the underlying physics processes, e.g.\ using multivariate techniques, can result in improved rejection of neutral pileup particles. For instance, some of the results shown in \cite{PUPPI} seem to suggest over-subtraction of soft QCD radiation, whereby particles originating from the hard parton scattering can be interpreted as pileup-related, and we expect the addition of a weighting algorithm not based on particle proximity to result in improved background rejection as the average pileup rate increases.

\subsection{Reshaping the particle-level kinematic distributions\label{fingerprints}}

In this section, we illustrate a different use of the particle weights with reference to the quantity $w_0(\eta, p_T)$ discussed in section \ref{weights}. As opposed to employing the weights to rescale the particle four-momentum vectors \cite{PUPPI,SoftKiller,berta}, we use them to reshape the particle-level $(\eta, p_T)$ distribution in each event. We show that this approach allows the estimation of the number of neutral soft QCD particles in each $(\eta, p_T)$ bin with reasonable accuracy regardless of whether or not signal particles are present.

Given an event and the definition of $w_0(\eta, p_T)$, the expected number of neutral soft QCD particles in each $(\eta, p_T)$ bin is given by:

\begin{equation}
\nu_b(\eta, p_T) = w_0(\eta, p_T) n(\eta, p_T),
\label{eq:sculpt}
\end{equation}

where $n(\eta, p_T)$ is the corresponding number of neutral particles in the data. Given $\nu_b(\eta,p_T)$, the unknown number of neutral soft QCD particles in each $(\eta, p_T)$ bin can be treated as a binomial random variable with mean given by (\ref{eq:sculpt}) and standard deviation:

\begin{equation}
\sigma_{\hat{n}_b}(\eta, p_T) = \left\{n(\eta, p_T) w_0(\eta, p_T) \left[1-w_0(\eta, p_T)\right]\right\}^{\frac{1}{2}}.
\label{eq:dnsculpt}
\end{equation}

In conclusion, the number of neutral soft QCD particles in each bin can be estimated in terms of:

\begin{equation}
\hat{n}_b(\eta, p_T) = w_0(\eta, p_T) n(\eta, p_T) \pm \sigma_{\hat{n}_b}(\eta, p_T).
\end{equation}

The quantity $\sigma_{\hat{n}_b}(\eta, p_T)/\hat{n}_b(\eta, p_T)$ is shown in figure \ref{fig:heatmap} (b) corresponding to one of the events analysed in this study. The performance of the algorithm is illustrated in figure \ref{fig:truth}, which provides a comparison between the true (a) and the estimated (b) number of neutral pileup particles across the $(\eta, p_T)$ space in the same event. The accuracy of the estimated shape of the distribution of neutral pileup particles in the event is further illustrated in figure \ref{fig:rats}, where the ratio to the true distribution of the control sample (a-d) and of the estimated distribution (e-h) is displayed as a function of particle $\eta$ in different $p_T$ bins in the region $0<p_T<0.2$~GeV/c. The error bars in (e-h) correspond to one binomial standard deviation on the bin contents in $\hat{n}_b(\eta, p_T)$.

It should be noted that the idea of employing the weights to reshape the $(\eta, p_T)$ distribution in the data can be used in conjunction with any definition of the weights, and that, in particular, it does not require all particles in the same $(\eta, p_T)$ bin to have equal weights. In fact, if $S(\eta, p_T)$ denotes the set of particles $i$ in the event inside a given $(\eta, p_T)$ bin, the procedure outlined above is equivalent to estimating $\hat{n}_b(\eta, p_T)$ according to $\hat{n}_b(\eta, p_T) = \sum_{i\in S(\eta, p_T)}w_i$, where $w_i$ is the weight assigned to particle $i$. The quantity $\sum_{i\in S(\eta, p_T)}w_i$ in fact reduces to $w_0(\eta, p_T) n(\eta, p_T)$ when all particles in the same $(\eta, p_T)$ bin have the same weight, $w_i = w_0(\eta, p_T)$. 

It is also worth noticing that the algorithm is inherently parallel, since different bins can be processed independently. We believe that the simplicity and parallelisation potential of this technique make it a suitable candidate for inclusion in future particle-level event filtering procedures upsteam of jet reconstruction at high-luminosity hadron collider experiments.




\section{Conclusions\label{concl}}

With reference to the upcoming higher-luminosity regimes of operation of the Large Hadron Collider, it is our opinion that the combination of different sets of particle weights encoding complementary information about the underlying physics processes has the potential to improve further on pileup subtraction, i.e.\ on the rejection of background particles originating from other proton-proton collisions. 

We have discussed a choice of weights that, unlike that employed by other methods, is not based on particle-to-particle proximity, but rather on the particle-level kinematic signatures of the signal hard parton scattering and of the low-energy strong interactions. We have also shown that, when the weights are used to reshape the particle-level kinematic distributions inside individual collision events, they lead to reasonable estimates of the number density of background neutral particles across the kinematic space. 

As more particle weighting methods become available, we envisage the possibility of developing algorithms based on multivariate combinations of different sets of weights with a view to exploiting all the particle-level information available in the data to reject neutral pileup particles. This study is based on a deterministic variant of a Markov Chain Monte Carlo algorithm that we previously discussed in conjunction with the idea of filtering individual collision events on a particle-by-particle basis at the reconstruction level in high-luminosity hadron collider environments. The main advantages of this approach, as compared to the previous stochastic version, are its parallelisation potential and the simplicity of the calculations involved. 


~\\
\section{Acknowledgments}
The author wishes to thank the High Energy Physics Group at Brunel University London for a stimulating environment, and particularly Prof. Akram Khan, Prof. Peter Hobson and Dr. Paul Kyberd for fruitful conversations, as well as Dr. Ivan Reid for help with technical issues. Particular gratitude also goes to people the author had fruitful discussions with at an early stage of development of this research idea, namely Prof. Jonathan Butterworth, Prof. Trevor Sweeting and Dr. Alexandros Beskos at University College London, as well as Prof. Carsten Peterson and Prof. Leif Lönnblad at Lund University.\\


\end{multicols}
\end{document}